\documentclass[12pt,a4paper,final]{iopart}

\usepackage{iopams}  
\usepackage[breaklinks=true,colorlinks=true,linkcolor=blue,urlcolor=blue,citecolor=blue]{hyperref}
\usepackage{cite}
\usepackage{graphicx}
\usepackage{caption}
\usepackage{epsfig}
\usepackage{epstopdf}
\usepackage{textcomp}
\usepackage{lineno}
\usepackage{ws-rotating} 

\begin{document}

\title[S.L. da SILVA et al.]{High Resolution Parameter Space from a Two Level Model on Semi-Insulating GaAs}



\author{S.L. da Silva$^{a, }$$^b$$^{, 1, }$; E.R. Viana$^{a, }$$^c$$^{, 2}$; A.G. de Oliveira$^{a}$;\\
G.M. Ribeiro$^{a}$ and  R.L. da Silva$^{d}$}
\vspace{1pc}
\address{$^{a}$Departamento de F\'{i}sica, Instituto de Ci\^{e}ncias Exatas, Universidade Federal de Minas Gerais, Caixa Postal 702, 30123-970 Belo Horizonte, Brazil.\\
\vspace{0.5pc}
$^{b}$Instituto Federal do Esp\'{i}rito Santo, campus Vit\'{o}ria, 29040-780 Vit\'{o}ria, Brazil.\\
\vspace{0.5pc}
$^{c}$Departamento de F\'{i}sica, Universidade Tecnol\'{o}gica Federal do Paran\'{a}, 80230-901 Curitiba, Brazil.\\
\vspace{0.5pc}
$^{d}$Instituto Federal Fluminense, campus Bom Jesus do Itabapoana, 28360-000 Bom Jesus do Itabapoana, Brazil.\\
\vspace{0.5pc}
\center{$^{1}$samir.lacerda@ifes.edu.br\\
$^{2}$emilsonfisica@ufmg.br}}
\vspace{2pc}

\begin{abstract}
Semi-insulating Gallium Arsenide (SI-GaAs) samples experimentally show, under high electric fields and even at room temperature, negative differential conductivity in N-shaped form (NNDC).Since the most consolidated model for n-GaAs, namely, ``the   model", proposed by E. Sch\"{o}ll was not capable to generate the NNDC curve for SI-GaAs, in this work we proposed an alternative model. The model proposed, ``the two-valley model" is based on the minimal set of generation-recombination equations for two valleys inside of the conduction band, and an equation for the drift velocity as a function of the applied electric field, that covers the physical properties of the nonlinear electrical conduction of the SI-GaAs system. The ``two-valley model" was capable to generate theoretically the NNDC region for the first time, and with that, we were able to build a high resolution parameter-space of the periodicity (PSP) using a Periodicity-Detection (PD) routine. In the parameter space were observed self-organized periodic structures immersed in chaotic regions. The complex regions are presented in a "shrimp" shape rotated around a focal point, which forms in large-scale a ``snail shell" shape, with intricate connections between different ``shrimps". The knowledge of detailed information on parameter spaces is crucial to localize wide regions of smooth and continuous chaos.
\end{abstract}

\vspace{2pc}

\section{Introduction}
\noindent 

In nonlinear dynamics, semiconductors behave as dissipative systems and can show, when placed under controlled conditions, a nonlinear electrical transport regime. A classical example is the semi-insulating Gallium Arsenide (SI-GaAs) material, which have nonlinear characteristic curves of the longitudinal current density as a function of the applied electric field $\epsilon$, {J($\epsilon$)}, with negative differential conductivity in a N-shaped form (NNDC),when high electric fields are applied, even at room temperature.

In the NNDC curves of SI-GaAs, it is possible to observe experimentally spontaneous electrical field oscillations namely the low-current oscillations (LFO) \cite{DaSilvaRL04, Rubinger03, Albuquerque03}. The time series of the LFO's, can be analyzed using nonlinear tools in order to create a bifurcation map where self-organized patterns are observed. It is important to note that the LFO´s are associated with the NDC curve of the material, and not to an oscillation of a circuit or device, as in the case of the Gunn´s diode \cite{Aoki01,Gunn63,Ridley63} and Chua's Diode \cite{Viana10}.

Other materials such as n-type GaAs, at cryogenic temperatures (e.g., 4.2K), shows the S-shaped form for NDC (SNDC) induced by the phenomena of impact ionization of shallow donors and field enhanced trapping \cite{Scholl86,Scholl87,Scholl01,Albuquerque05,Tzeng09}. The most consolidated model to explain this behavior in materials with shallow-level is the well known ``two-level model", as proposed by E. Sch\"{o}ll \cite{Scholl86,Scholl87,Scholl01,Aoki01}. In Sch\"{o}ll's model the conduction band (CB), the fundamental level of the defect (1s-ground state at 5.9 meV below the (CB) and its first excited level (2p-excited state at 1.5 meV below the (CB) are the energy-levels responsible for the nonlinear dynamics. Self-organized patterns arise from the generation-recombination (g-r) rate equations as a function of the applied electric field \cite{Scholl86,Albuquerque05,Tzeng09,Albuquerque06}.

All studied models until now based on g-r rate equations, not used the NNDC curve to explain the experimental results in SI-GaAs, only the SNDC characteristic. In order to cope with this problem, in this work we had proposed an alternative physical model for the conduction phenomena in SI-GaAs that successfully generates the NNDC curve and consequently the nonlinear dynamics. The two-levels inside the band-gap of the Sch\"{o}ll´s model were replaced by two valleys, namely the  $\Gamma$ and L-valley inside the conduction band and we named our model as the "two-valley model". 

In the ``two-valley model" the first excited state of the defect was eliminated, the fundamental level of the deep-defect was preserved, assuming as deep level, and the conduction mechanisms involving the two valleys \cite{Blakemore82,Yu01} of the GaAs were introduced. This new model proposed by us is similar to the intervalley transfer of the Gunn effect \cite{Aoki01,Gunn63,Ridley63}, but the energy levels involved, and the dynamics are very different, since in the Gunn effect, we are dealing with the regime of high frequency oscillations (GHz range) and at very low temperatures (T  ${<}$ 10K). Besides that, the models proposed to explain the Gunn effect were not designed to study the nonlinear dynamics, they are only used to explain the high frequency oscillations that emerge from this device and not to study their physical properties \cite{Hidetaka01,Cenys92}.

By solving the g-r rate equations for our model we successfully produced the NNDC region in the {J($\epsilon$)} curves and the nonlinear dynamics are present, as spontaneously low frequency (sub-Hz to about sub-GHz) current oscillations, the LFO's.

In this paper the phenomenology of the nonlinear NNDC curve will be evaluated in detail, by a group of different attractors, and their correspondent bifurcation diagram. We have also analyzed the time series using a standard Periodicity-Detection (PD) routine \cite{Viana12}, implemented in LabVIEW, and the Parameter-Space of the periodicity (bifurcation diagram codimension-two) was built with the reconstructed attractors.

Within the ``two-valley model" we now have an appropriate model to study the nonlinear electrical conduction of the SI-GaAs system that has been doubt for the past years, since until this time, no model could generate the NNDC curve of SI-GaAs. The nonlinear conduction of SI-GaAs includes periodic and chaotic regimes, bifurcation routes to chaos, periodic structures immersed in chaotic regions, as those observed experimentally \cite{DaSilvaRL04, Rubinger03, Albuquerque03}.

Although the bifurcation phenomena involving only one control parameter (codimension-one) are reasonably understood today, when we increase the number of control parameters, we experience the emergence of complex codimension-two periodic regions immersed in chaotic regions forming self-similar structures, as shrimps \cite{Viana10,Hoff14,Bonatto08,Albuquerque10,Stoop10,Oliveira11}, that still needed a full theoretical and experimental description. The knowledge of this complex periodic region is important to fulfill the requirements for secure communications with chaos \cite{Baptista98}, since with a periodic-attractor no information can be encrypted for communication.

In this work, we proposed a model, ``the two-valley model", that are based on rate equations with the minimal set of generation-recombination equations for two valleys inside of the conduction band, and an equation for the drift velocity as a function of the applied electric field, which considers a conduction mechanism involving the two valleys on conduction band and the fundamental level of the deep-defect, present in section 2. In section 3, our numerical results are capable to generate theoretically the NNDC region, and in section 4 the nonlinear dynamics will be investigated through the attractors built by a combination of the time series of the defect density and the electric field. Many periodic and chaotic regimes are observed in the bifurcation diagram as well as a period-doubling bifurcation route-to-chaos with odd-periodic windows. In the same section, we will be build a high resolution parameter-space of the periodicity (PSP) using a Periodicity-Detection (PD) routine. In the parameter space were observed self-organized periodic structures immersed in chaotic regions.

\section{The two-valley model}
Experimentally, it is well known that under high electric fields the drift velocity of free electrons in GaAs is nonlinear because it involves mixed conduction in $\Gamma$ and L valleys with different mobility values and with  $\mu_{\Gamma}$ ${>}$ $\mu_{L}$ \cite{Blakemore82,Yu01}.

\begin{figure}[h]
\begin{center}
\includegraphics{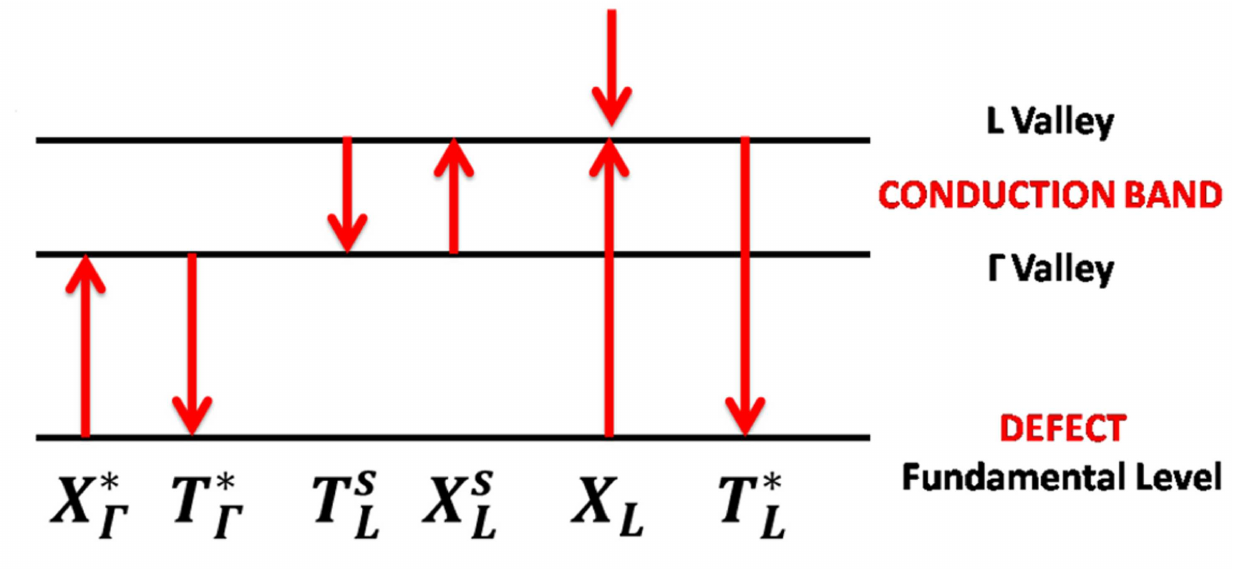}
\end{center}
\caption{Generation-recombination (g-r) processes, involving the $\Gamma$ and the L valleys, inside the conduction band and the fundamental state of the deep defect (a middle-gap level with $\Delta E_{band-gap}$=1.42eV at room temperature), that was considered in the two-valley in SI-GaAs.}
\label{Figure1}
\end{figure}

The two-valley model, proposed by us, includes the fundamental level of the deep defect, and the $\Gamma$ and the L valleys inside the conduction band, with $\Delta$E$_{\Gamma-L}$ = 290 meV \cite{Blakemore82}. The valence band is omitted because the recombination rates involving it are negligible. The carrier density $\nu$ in the conduction band is governed by the g-r process in which an electron in the ground state of the impurity may be thermally or impact ionized to the conduction band and may then recombine a donor having an empty state. Inside the conduction band the g-r process involves mixed conduction in $\Gamma$ and L valleys. In Fig.~\ref{Figure1}, we present the schematic diagram of the g-r rate equations that we have used in order to generate the N-shaped curve from the two-valley model. The up arrows represent the generation and the down arrows, the recombination processes of the electrons. In the present model the coefficients of generation processes (X$_{L}$ - impact ionization; X$_{L}^{S}$ - intervalley  process), the coefficients of recombination processes (T$_{L}^{*}$ - field enhanced trapping; T$_{L}^{S}$ - intervalley process), depend on the electric field. All other coefficients (X$_{\Gamma}^{*}$ and T$_{\Gamma}^{*}$) were constant and represent the coefficients for transitions from the fundamental state of the deep level defect to $\Gamma$ valley state and vice versa, respectively.  

The electron density in the valley $\Gamma$ $\nu$$_{\Gamma}$ and valley L $\nu$$_{L}$ in the conduction band, the fundamental state of the deep level defect $\nu$$_{D}$ can be expressed through the g-r rate equations:

\begin{equation}
\dot{\nu}_k = \varphi_k(\nu_\Gamma,\nu_L,\nu_D, E), k=index(\Gamma,L)
.\label{eq001}
\end{equation}
\begin{equation}
\dot{\nu}_D = \varphi_D(\nu_\Gamma,\nu_L,\nu_D, E)
.\label{eq002}
\end{equation}

\vspace*{2mm}
\noindent
with E the electric field, and the dot denoting the time derivative. The concentrations and fields are coupled via Maxwell’s equations for the charge density $\epsilon_S \nabla\cdot E = e(N_D^*-\varphi_{D}-\varphi_{L}-\varphi_{\Gamma})$, where $N_D^*$ is the effective donor density and $\epsilon_S$ is the dielectric constant. The $\varphi_{k}$ and $\varphi_{D}$ are the generation-recombination rates which are - for nondegenerate statistics - polynomial functions of the concentrations. They depend implicitly upon the electric field through the rate constants, e.g. impact ionization coefficient, field enhanced trapping coefficient and coefficients involving the $\Gamma$ and L valleys. Since the total number of electrons is conserved by g-r processes, $\varphi_{\Gamma}$+$\varphi_{L}$+$\varphi_{D}$= 0 holds \cite{Scholl86, Scholl87}.

Using normalized variables \cite{Scholl86, Scholl87} for the charge density of the fundamental state of the deep level defect, n$_{D}$, the charge density in the $\Gamma$ valley, n$_{\Gamma}$, and those respective derivatives, and  the dimensionless time $\tau$, length $\xi$, and $\epsilon$ electrical field variables, we have:

\begin{equation}
n_i = (\nu_i/N_D^*), i=index(D,L,\Gamma)
.\label{eq01}
\end{equation}
\begin{equation}
\tau \equiv (t/\tau_M), \tau_M \equiv (\epsilon_S/4 \pi \mu_0 N_D^*)
.\label{eq02}
\end{equation}
\begin{equation}
\xi \equiv (\vec x/L_D), L_D \equiv (D_0 \tau_M)^{1/2}
.\label{eq03}
\end{equation}
\begin{equation}
\vec \epsilon \equiv (e L_D/k_B T) \vec E
.\label{eq04}
\end{equation}
\vspace*{2mm}

\noindent 
where $N_D^*$ is the effective donor density, $\tau_M$ is the effective relaxation time, $\epsilon_S$ dielectric constant, $L_D$ Debye effective length, $D_0$ diffusion constant, \textit{e} electron charge, $k_{B}$ Boltzmann’s constant, T temperature and  $\mu_0$ mobility. So, the normalized set of g-r rate equations represented in Fig.~\ref{Figure1} is given by:

\begin{equation}
\dot{n}_\Gamma =X_\Gamma^* n_D -T_\Gamma^* n_\Gamma p +T_L^S n_L -X_L^S n_\Gamma
.\label{eq05}
\end{equation}
\begin{equation}
\dot{n}_D =-X_\Gamma^* n_D +T_\Gamma^* n_\Gamma p -X_L n_D n_L +T_L^* n_L p
.\label{eq06}
\end{equation}
\begin{equation}
\dot{\epsilon} =J(\epsilon)-n_\Gamma V(\epsilon)
.\label{eq07}
\end{equation}

\vspace*{2mm}
The effective density of donors and acceptors are represented by $N_D^*$ and $N_A$, respectively, and the charge density in the L valley $n_L$ was eliminated by the charge neutrality condition, namely $n_L=(N_D^* -n_D -n_\Gamma)$. The density p of the unoccupied state of the deep defect is given by $p=(N_A+n_\Gamma+n_L)$.

The g-r transition coefficients between the fundamental state and the $\Gamma$ valley are $X_\Gamma^*$ and $T_\Gamma^*$, respectively. Both coefficients depend on the light intensity and temperature, and they control the relative charge densities in the conduction channels, and they are supposed to be independent of $\epsilon$.

The generation coefficient $X_L$ stands for the impact ionization process involving the fundamental state and the valley L, and $T_L^*$ stands for the phenomenon of field enhanced trapping \cite{Rubinger00}. Their functional dependence with the applied electrical field $\epsilon$ is given by:

\begin{equation}
X_L =X_L^0 exp(-E_{XL}/\epsilon)
.\label{eq08}
\end{equation}
\begin{equation}
T_L^* =T_L^{0*} exp(-E_{TL*}/\epsilon)
.\label{eq09}
\end{equation}

\vspace*{2mm}
The g-r coefficients and involving the $\Gamma$ and L valleys, with an electric field dependence $\epsilon$, is giving by:

\begin{equation}
X_L^S =X_L^{S0} exp(-E_{SL}/\epsilon)
.\label{eq10}
\end{equation}
\begin{equation}
T_L^S =T_L^{S0} exp(-E_{S\Gamma}/\epsilon)
.\label{eq11}
\end{equation}

\vspace*{2mm}
The conduction mechanism involving the two valleys may depend on some parameters such as temperature, light intensity, electric and magnetic fields. However, in this work, we only deal with the dependence of the electric field $\epsilon$ since been the easiest parameter to be controlled experimentally, by applying an electrical voltage in the system.

It is possible to incorporate as many other mechanisms, as one wishes, into the present model, but they may become superfluous and therefore have very little effect on the final results, because the g-r rate equation chosen, eqs.~(\ref{eq08}-\ref{eq11}) are the smallest set tested that still produces the NNDC region, and take into account the main SI-GaAs physical properties \cite{DaSilvaRL04,Rubinger03,Albuquerque03,DaSilvaRL06}. Furthermore, for simulations the smallest set of equations that still reproduces all physical phenomena is important, in order to reduce the computational cost.

The eqs.~(\ref{eq08}-\ref{eq11}) have dependence with the electric field $\epsilon$ consolidated in the literature, by the Shockley's model \cite{Schockley61}, where $E_i$ are critical fields, related to the activation energies, since the ionizing carrier must have a certain threshold energy before it can ionize or captured. The constants preceding the exponential term are the upper limit for the generation (X$_{L}^{0}$, X$_{L}^{s0}$) and recombination (T$_{L}^{*0}$, T$_{L}^{s0}$) rates. The field-enhanced trapping process originates from a fully nonradiative relaxation multiple phonon emission capture process by the As antiside defects assisted by the applied electrical field \cite{Rubinger00}. It’s assumed to follow the same model except by X$_{TL*}$ is the threshold capture energy. The g-r equations, eqs.~(\ref{eq08}-\ref{eq11}), have a low dependence of $\epsilon$ for low electric fields, and with an increase of $\epsilon$, we should have high dependence of the rates with the electric field. 

In Sch\"{o}ll model \cite{Scholl86,Scholl87,Scholl01,Aoki01}, the drift velocity curve was modeled by the empirical saturable form arctangent curve which increases linearly with the electrical field for small $\epsilon$ and saturates for large $\epsilon$, which is an approximation of the curve of the GaAs. In our model ``two-valley model" provides an inter-play mechanism between the valleys, where the transferred-electron effect is the transfer of conduction electrons from a high-mobility energy valley $\Gamma$ to low-mobility, higher-energy satellite valleys L. This effect leads to negative differential resistance on drift-velocity curve \cite{Yu01}. This model is more appropriate than saturable form arctangent curve to be a drift-velocity curve that shows a competition between valleys, which is more realistic curve for GaAs. This is the crucial role that the ``two-valley model" plays in the N-type current density-electric field characteristics for SI-GaAs.

So the important point of our model concerns the drift-velocity behavior. It implies a change from the empirical saturated form of an arctangent as function of $\epsilon$, as proposed by Sch\"{o}ll in the two-level model. This change has its physical origin in the mixed conduction of the free-electrons in the $\Gamma$ and L valleys, as a function of $\epsilon$. Indeed, in order to get the new effective drift velocity V($\epsilon$) and to test our model, we have used the experimental data of V($\epsilon$) for GaAs \cite{Ruch68}, and we have modeled it by:

\begin{equation}
V(\epsilon) =(\mu_0+\mu_1e^{(-E_1/\epsilon)}-\mu_2e^{(-E_2/\epsilon)})\epsilon
.\label{eq12}
\end{equation}

\vspace*{2mm}
\noindent 
where $\mu_0$, $\mu_1$ and $\mu_2$ are constants with a dimension of electric mobility, $E_1$ and $E_2$ are critical electrical fields. 

\begin{figure}[h]
\begin{center}
\includegraphics{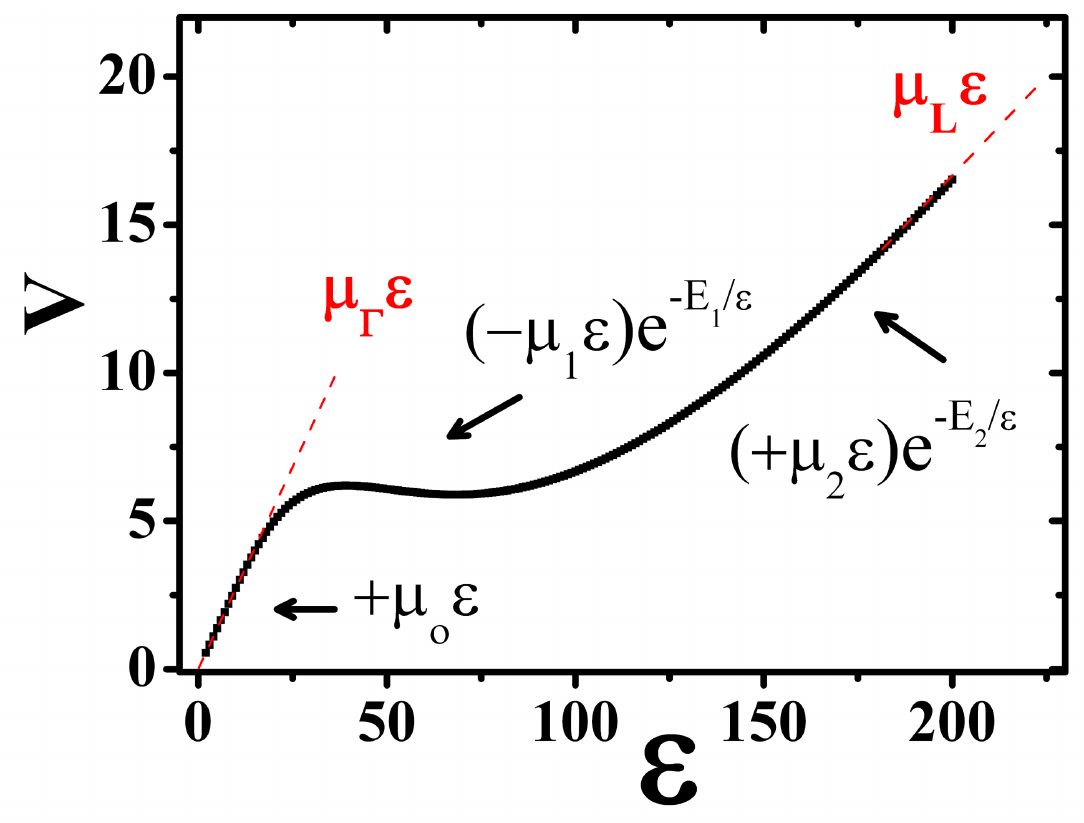}
\end{center}
\caption{The effective drift velocity V($\epsilon$) of the free electrons in SI-GaAs, experimental data of \cite{Ruch68} that was modeled using eq.~(\ref{eq12}). For low values of $\epsilon$, the linear term ($+\mu_0 \epsilon$) is dominant with $\mu_0  \rightarrow \mu_\Gamma$. For intermediate values of $\epsilon$, the negative term (-$\mu_1 \epsilon$)$e^{(-E_1/\epsilon)}$ becomes the effective one and produces a decreasing in V($\epsilon$) which implies in negative slope. Finally, for high values of $\epsilon$, the positive term (+$\mu_2 \epsilon$)$e^{(-E_2/\epsilon)}$ becomes the dominant one with a linear positive slope and with $\mu_2  \rightarrow \mu_L$.}
\label{Figure2}
\end{figure}

For low values of $\epsilon$, the linear term ($+\mu_0 \epsilon$) is dominant with $\mu_0  \rightarrow \mu_\Gamma$. For intermediate values of $\epsilon$, the negative term (-$\mu_1 \epsilon$)$e^{(-E_1/\epsilon)}$ becomes the effective one and produces a decreasing in V($\epsilon$) which implies in negative slope. Finally, for high values of $\epsilon$, the positive term (+$\mu_2 \epsilon$)$e^{(-E_2/\epsilon)}$ becomes the dominant one with a linear positive slope and with $\mu_2  \rightarrow \mu_L$ as shown in Fig.~\ref{Figure2}.

In order to get the curve {J($\epsilon$)}, we need to find the steady state regime ($\dot{n}_\Gamma =\dot{n}_D =\dot{\epsilon}= $0) in the eqs.~(\ref{eq05}-\ref{eq07}) producing a polynomial of third order degree, $P_3$($n_\Gamma$)=0. Similar equations can be obtained from $n_L$. Each of the three zeros for $n_\Gamma (\epsilon)$, namely $n_{\Gamma 1}$, $n_{\Gamma 2}$ and $n_{\Gamma 3}$, produces an independent  {$J_i (\epsilon)$} curve, using the following equation:

\begin{equation}
J_i (\epsilon) =n_{\Gamma i}(\epsilon)V(\epsilon) 
.\label{eq13}
\end{equation}

\vspace*{2mm}
For physical reasons only one of the three independent {$J_i (\epsilon)$} curves is a solution, because when one curve is stable, the others become unstable.

The two-valley and two-level models are different. The two-level model is applied to n-GaAs, at cryogenic temperatures (e.g., 4.2K) and has energy levels of shallow defects. its g-r rate equations shows the S-shaped form curve (SNDC)  obtained for a speed of drift of the arc tangent form . The  two-level model does not consider conduction channels in two valleys on conduction band, only the conduction band, the fundamental level of the defect and its first excited level. The g-r rate depend on the electric field is represented by impact ionization of shallow donors \cite{Scholl86,Scholl87,Scholl01,Albuquerque05,Tzeng09,Albuquerque06}. In [Albuquerque, 2006], the authors introduce into the two-level model the recombination process of ``field enhanced trapping" for explain this nonlinear phenomenon in SI-GaAs sample. But, this model do not generate the NNDC curve, as presented experimentally for SI-GaAs \cite{DaSilvaRL04, Rubinger03, Albuquerque03}.

The ``two-valley model" is applied to SI-GaAs, at room temperature, with deep-defect energy level and its g-r rate equations shows N-shaped form curve (NNDC) obtained for an analytical expression for the drift velocity - eq.~(\ref{eq12}), which considers a conduction mechanism involving the two valleys on conduction band and the fundamental level of the deep-defect. Our model produces the NNDC curve and the nonlinear dynamics as observed experimentally \cite{DaSilvaRL04, Rubinger03, Albuquerque03}.

\section{Numerical results}

\begin{table}[ht]
 \caption{Dimensionless g-r coefficient values and the parameters used in the simulations of the ``two-valley model".}
 \begin{center}
{\begin{tabular}{c c c c c c}\hline \hline \\[-2pt] 

$X_\Gamma^*$ & $1.0\times10^{-5}$ &$X_L^{S0}$& $1.5\times10^{-3}$ & $\mu_0$ & 0.275\\[6pt]
\hline\\[-2pt]
$T_\Gamma^*$ & $2.0\times10^{-5}$ &$E_{SL}$& 2.5 & $\mu_1$ & 0.935\\[6pt]
\hline\\[-2pt]
$X_L^0$ & $1.0\times10^{-2}$ &$T_{L}^{S0}$& $5.25\times10^{-3}$ & $\mu_1$ & 0.850\\[6pt]
\hline\\[-2pt]
$E_{XL}$ & 10.5 &$E_{S\Gamma}$& 22.5 & $E_1$ & 70\\[6pt]
\hline\\[-2pt]
$T_{L}^{*0}$ & $1.0\times10^{-2}$ &$N_A/N_D^*$& 0.135 & $E_2$ & 120\\[6pt]
\hline\\[-2pt]
$E_{TL*}$ & 1.0 & $-$ & $-$ & $-$ & $-$\\[6pt]
\hline \hline
\end{tabular}}
\end{center}
\end{table}

In this section, we present the numerical integration of eqs.~(\ref{eq05}-\ref{eq07}) in the steady state regime for the two-valley model in SI-GaAs, where the electric field $\epsilon$ was used as the control parameter. The g-r rate equations are given by eqs.~(\ref{eq05}-\ref{eq07}), and the g-r coefficients by eqs.~(\ref{eq08}-\ref{eq11}) and the drift velocity by eq.~(\ref{eq12}). The characteristic curve {J($\epsilon$)} were calculated by using eq.~(\ref{eq13}).

\begin{figure}[h]
\begin{center}
\centering\includegraphics[scale=0.4]{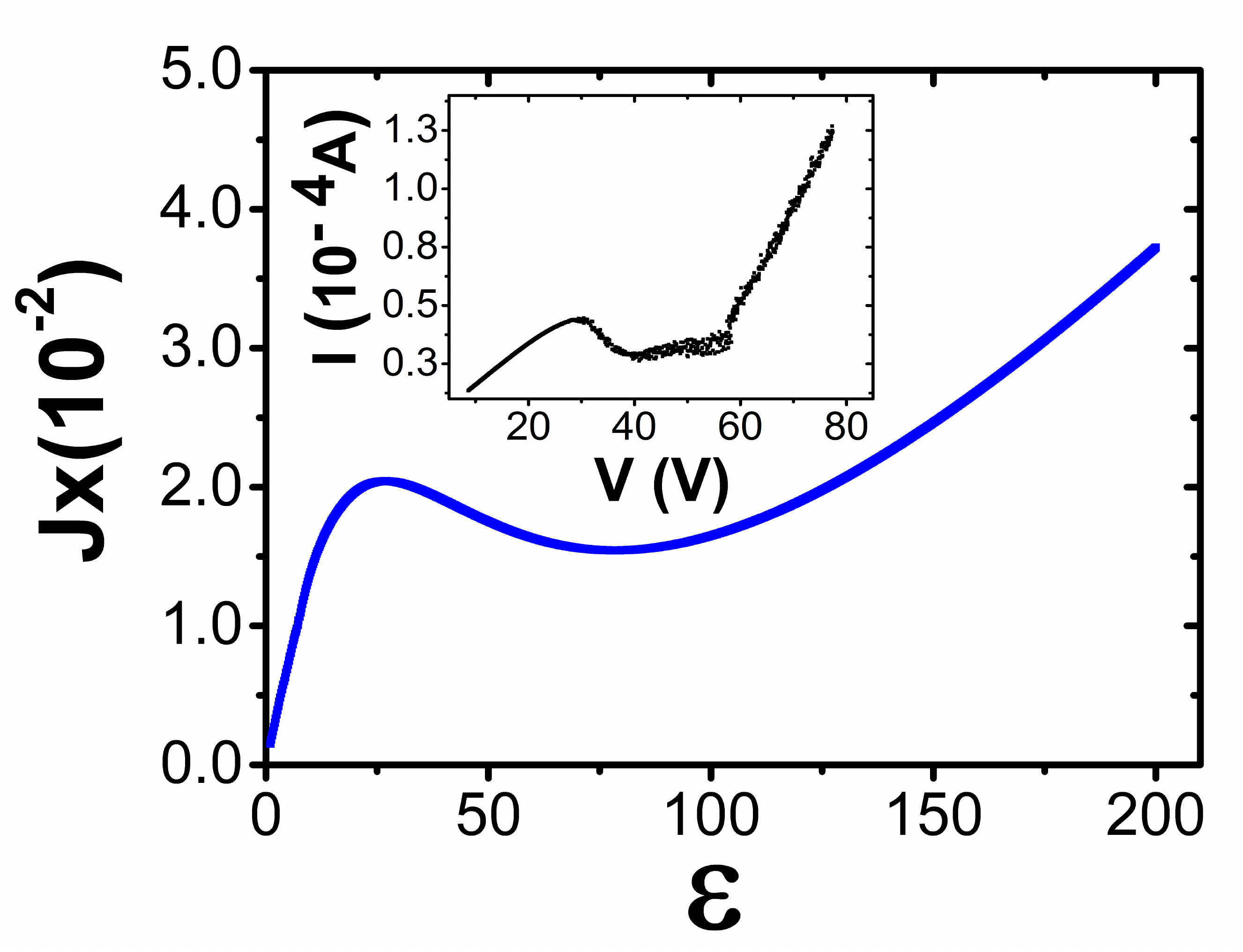}
\end{center}
\caption{The $\Gamma$ valley longitudinal current density {$J(\epsilon)$} for the two-valley model. In the inset, we show the NNDC region of the characteristic SI-GaAs experimental I(V) curve at 200K \cite{Rubinger03}.}
\label{Figure3}
\end{figure}

The stable {$J(\epsilon)$}  curve, presented in Fig.~\ref{Figure3}, clearly shows the NNDC characteristic curve, that were one of the objectives to be addressed in this work. In the inset of Fig.~\ref{Figure3}, we present the experimental curve I(V) showing the characteristics NNDC region at 200K, under illumination with a constant current of 30 mA through the LED, and for more details see \cite{Rubinger03}. In Table 1, we present the set of values used in our “two-valley model” that successfully generate the NNDC curves in Fig.~\ref{Figure3}.

\section{Nonlinear Time Series Analysis}

\subsection{Attractors and Bifurcation Diagrams}

We use the $4^{th}$ order Runge-Kutta method for numerical integration of the ``two-valley model" along with the parameter values presented in Table 1, in order to generated the time series $\epsilon(t)$, $n_\Gamma(t)$, and $n_D(t)$ for the SI-GaAs system. 

Each time series obtained contains all nonlinear information on the SI-GaAs system, since eqs.~(\ref{eq05}-\ref{eq07}) are coupled ordinary differential equations. For convenience, the time series analysis of $\epsilon(t)$ were carried in order to extract the nonlinear behavior of our proposed model, and to corroborate with the previous experimental results \cite{DaSilvaRL04, Rubinger03, Albuquerque03}. Where the electric field $\epsilon$ was used as the control parameter.

\begin{figure}[h]
\begin{center}
\includegraphics{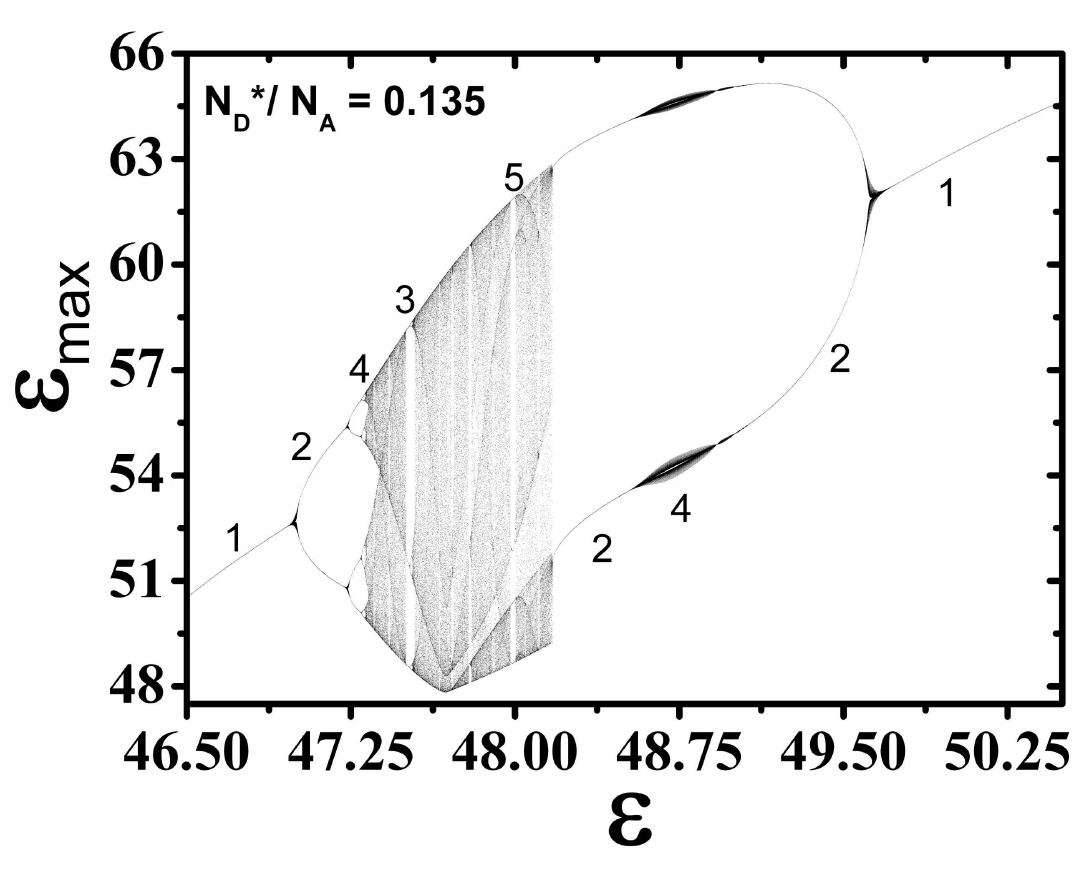}
\end{center}
\caption{Bifurcation diagram for the maximum of the electric field $\epsilon(t)$ time series obtained from the two-valley model in SI-GaAs.}
\label{Figure4}
\end{figure}

By using Poincar\'{e} maps, the bifurcation diagram of the maximum value of the electric field $\epsilon_{max}(t)$  was built and is presented in Fig.~\ref{Figure4}. In the bifurcation diagram, we show the well-known period-doubling bifurcation route-to-chaos: (period-1) - (period-2) - (period-4) - ${...}$ - (chaos). Besides that, is also possible to observe stable odd-periodic windows, as (period-3) and (period-5), embedded in the chaotic region, which was also observed experimentally for the SI-GaAs system \cite{DaSilvaRL04,DaSilvaSL09} and for other systems \cite{Viana10,Viana12}.

\begin{figure}[h]
\begin{center}
\includegraphics{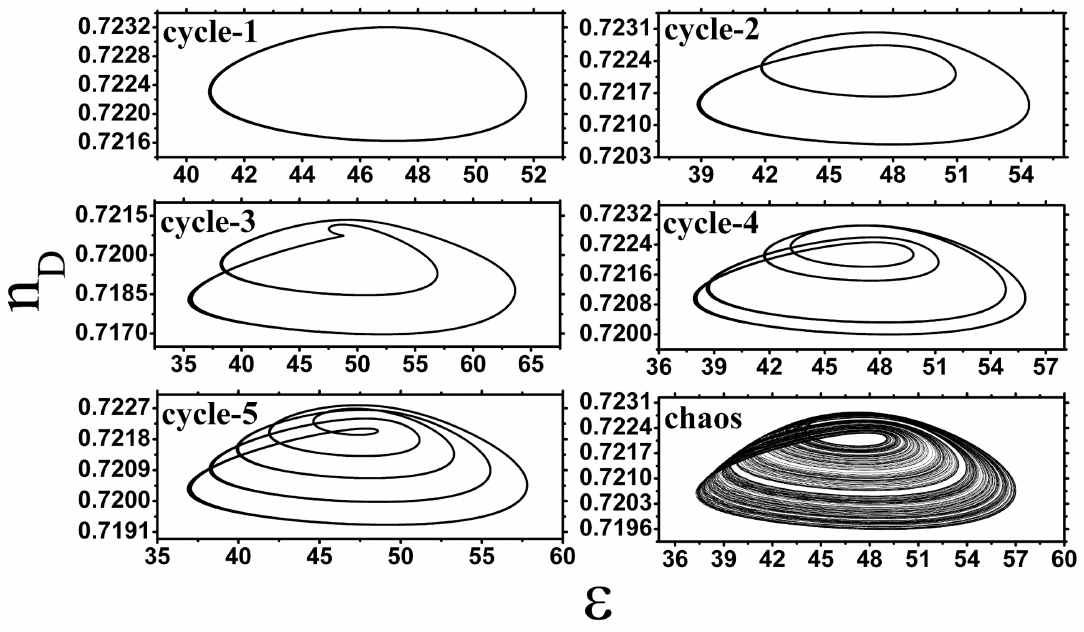}
\end{center}
\caption{Periodic attractors, (period-1) until (period-5), and a chaotic attractor, obtained from the two-valley model in SI-GaAs.}
\label{Figure5}
\end{figure}

In the Fig.~\ref{Figure5} periodic attractors from (period-1) to (period-5) and also chaotic attractors, were obtained, by using the time series of the defects density $n_D(t)$  and electric field $\epsilon$. 

\subsection{Periodicity Parameter-Space}

In order to explore the periodic structures that appear immersed in the chaotic regime and determine how they arise, we now shift our attention to a two-dimensional parameter-space, where two parameters are varied. In this case, we have used as the control parameters, the compensation factor $(N_A/N_D^*)$ and the electric field $\epsilon(t)$.

The compensation factor $(N_A/N_D^*)$ is only specified by the manufacturing process of the GaAs and is thus very difficult to control and measure experimentally. We would therefore like to describe the full configuration of the phase-space that emerges from the nonlinear transport of the SI-GaAs system. For each value of $(N_A/N_D^*)$ we have a bifurcation diagram as we change the electric field $\epsilon$, similar of the figure Fig.~\ref{Figure4}. If we want to determine the full phase-space, we have to change simultaneously the parameter $(N_A/N_D^*)$ and $\epsilon$, which is the emergence of the concept "Parameter-Space". 

The Periodicity Parameter-Space (P-parameter-space) studied in this work is a two-dimensional map ($(N_A/N_D^*)$,$(\epsilon)$) where the value of the periodicity P, of the time series $S(t)$, was encoded in a color scale, with 16 values. We have found in the literature some methods to determine the periodicity of the time series \cite{Hegger99,Hegger07}, but, in this work, we have used, for convenience, a fast and reliable Periodicity-Detector (PD) routine done in a LabVIEW environment, developed by \cite{Viana12}. The method consists in transforming the time-series $S(t)$ (flow) into a discrete time-series $U(n)$ (unitary) by selecting the maximum and minimum values of $S(t)$. Then, the number of peaks in the Fast Fourier Transformation (FFT) of $U(n)$ is counted, and this number defines the periodicity of the series $S(t)$. In this work, we count periods up to (period-16), so any series with P ${>}$ 16 was considered chaotic.

\begin{figure}[h]
\begin{center}
\centering\includegraphics[scale=0.4]{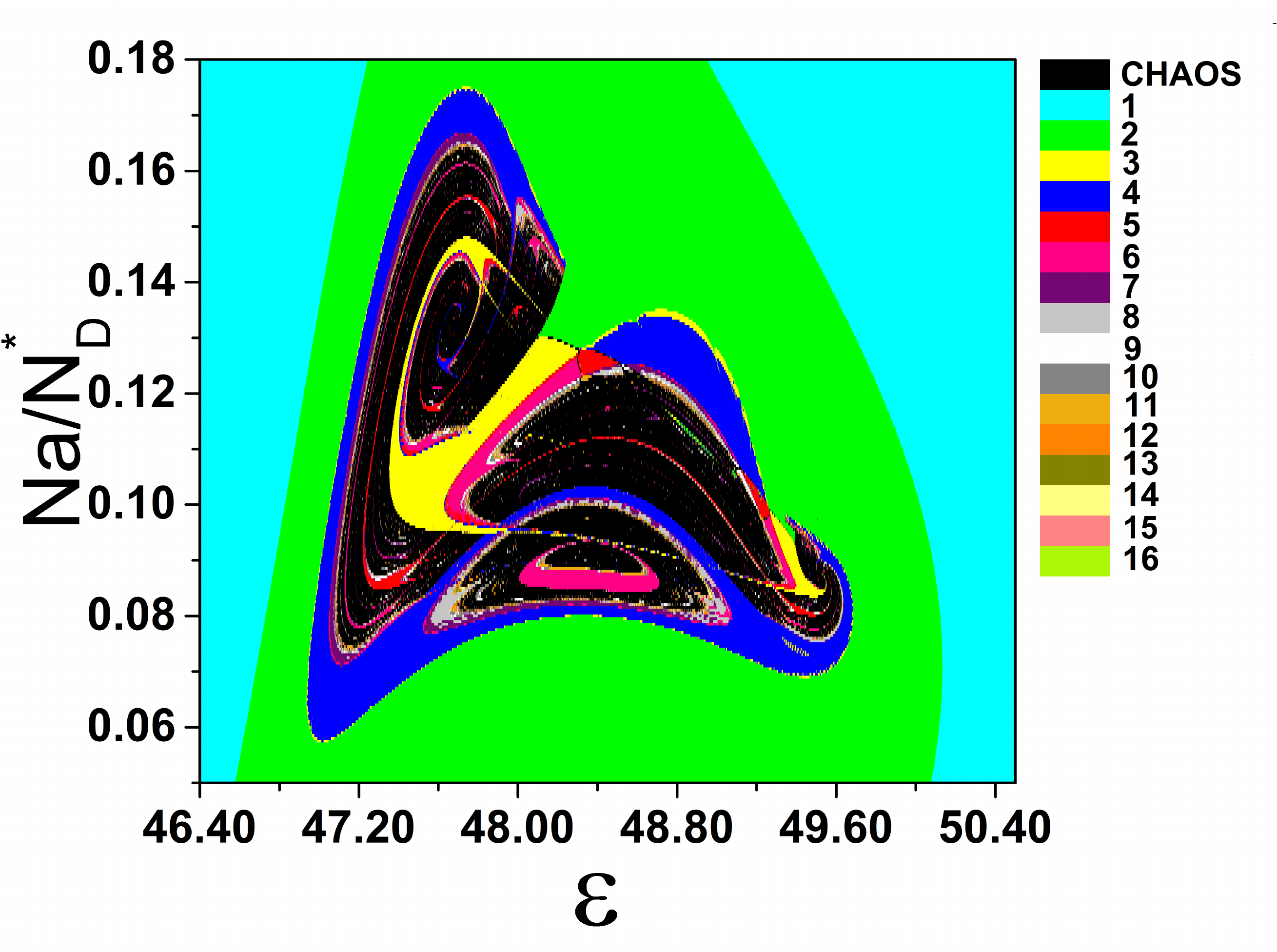}
\end{center}
\caption{(Color online) Parameter-space of the periodicity for two-valley model in SI GaAs. Here, compensation factor $(N_A/N_D^*)$ was varied from $5.0\times10^{-2}$ up to $18.0\times10^{-2}$ with $5.0\times10^{-4}$ step and the electric field $(\epsilon)$ from 46.4 up to 50.5 with $1.0\times10^{-3}$ step.}
\label{Figure6}
\end{figure}

\begin{figure}[h]
\begin{center}
\centering\includegraphics[scale=0.4]{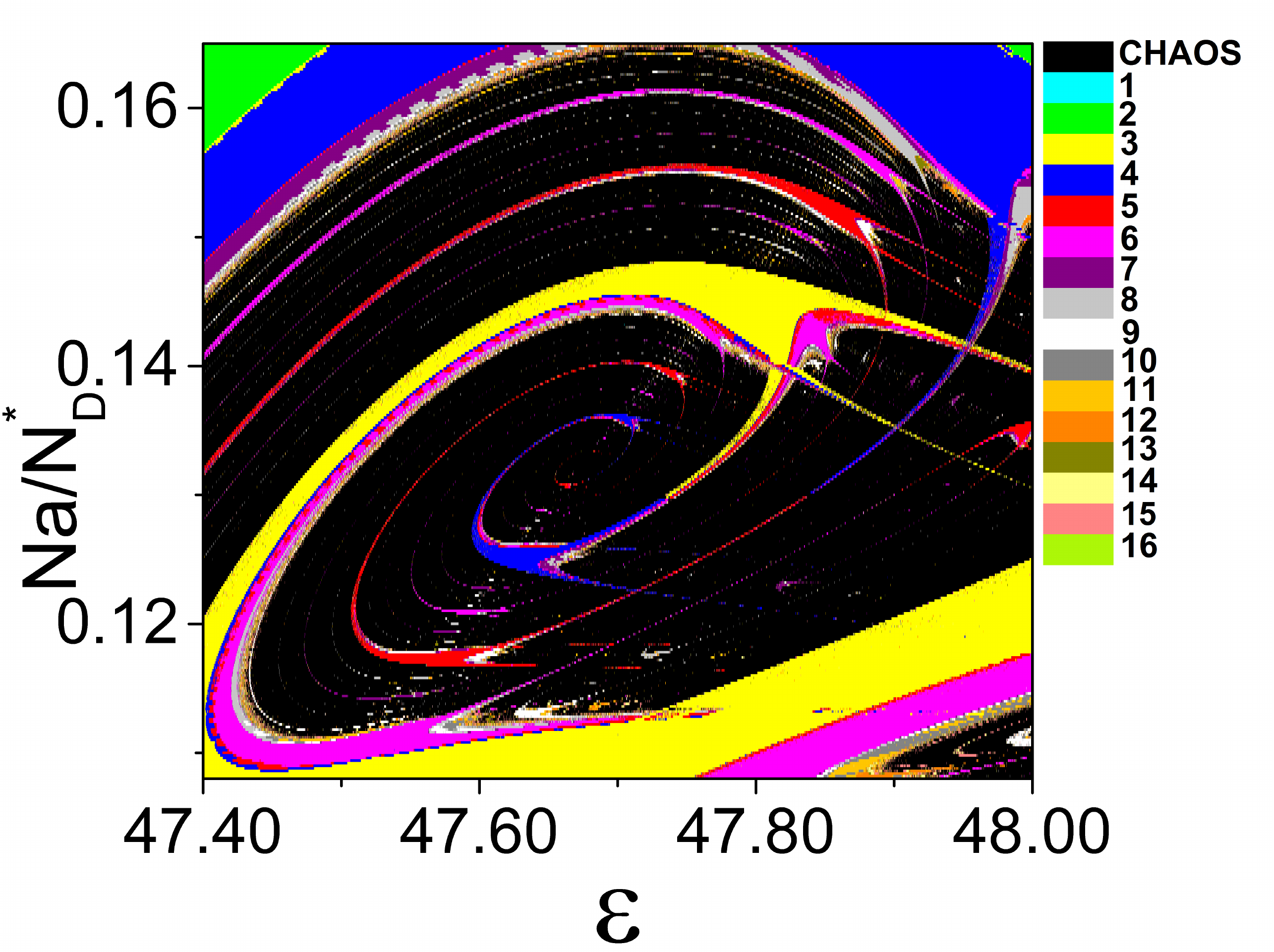}
\end{center}
\caption{(Color online) Magnification (25x) of the P-parameter-space for the SI-GaAs simulated by the ``Two-Valley Model". Here, compensation factor $(N_A/N_D^*)$ was varied from $10.8\times10^{-2}$ up to $65.0\times10^{-2}$ with $2.0\times10^{-4}$ step and the electric field $(\epsilon)$ from 47.4 up to 48.0 with $1.0\times10^{-4}$ step.}
\label{Figure7}
\end{figure}

In Fig.~\ref{Figure6}, we present our periodicity parameter-space of the SI-GaAs simulated using our ``two-valley model". The vertical axis, stands for the compensation factor $(N_A/N_D^*)$, from $5.0\times10^{-2}$ up to $18.0\times10^{-2}$ with $5.0\times10^{-4}$ step, and the abscissa stands for the electric field $(\epsilon)$ from 46.4 up to 50.5 with $1.0\times10^{-3}$ step. The parameter-space with a resolution $(x,y)$ = $(4100\times260)$ presented in Fig.~\ref{Figure6}, shows an interesting structural similarity to a ``snail", with a large structural characteristic of a "shrimps" of periodicity-3, embedded inside the ``snail shell".

The information on many periodic-structures, the ``complex periodic-windows", in the presented P-parameter-space now can be accessed. In some direction the periodic structures organize themselves in a period-adding bifurcation cascade, and the periodic regions are alternated by chaotic regions.

In order to get a better look of the "shrimps" in Fig.~\ref{Figure7}, we also show a magnification of (25 times) of one region of the parameter-space of the Fig.~\ref{Figure6}, with $(N_A/N_D^*)$ from $10.8\times10^{-2}$ up to $16.5\times10^{-2}$ with $2.0\times10^{-4}$ step, and $(\epsilon)$ varying  from 47.4 up to 48.0 with $1.0\times10^{-4}$ step, which implies in a resolution of $(6000\times280)$. Now, we can identify that the expanded parameter space show organized periodic structures known as ``shrimps", and they are oriented in a rotational form, around a focal point. So, the ``shrimps" organized in a spiral form, and in large-scale as the ``snail shell" on Fig.~\ref{Figure6}. Similar structures are already present in the literature for other systems \cite{Hoff14,Bonatto08,Albuquerque10,Stoop10}.

For better orientation, we describe the self-organized structure of the Parameter Space as the physical structure of a ``shrimp", where the body is composed by a large periodic regions having on the right, ramifications, that define ``antennas" and, on the left, the ``tail". 

\begin{figure}[h]
\begin{center}
\centering\includegraphics[scale=0.4]{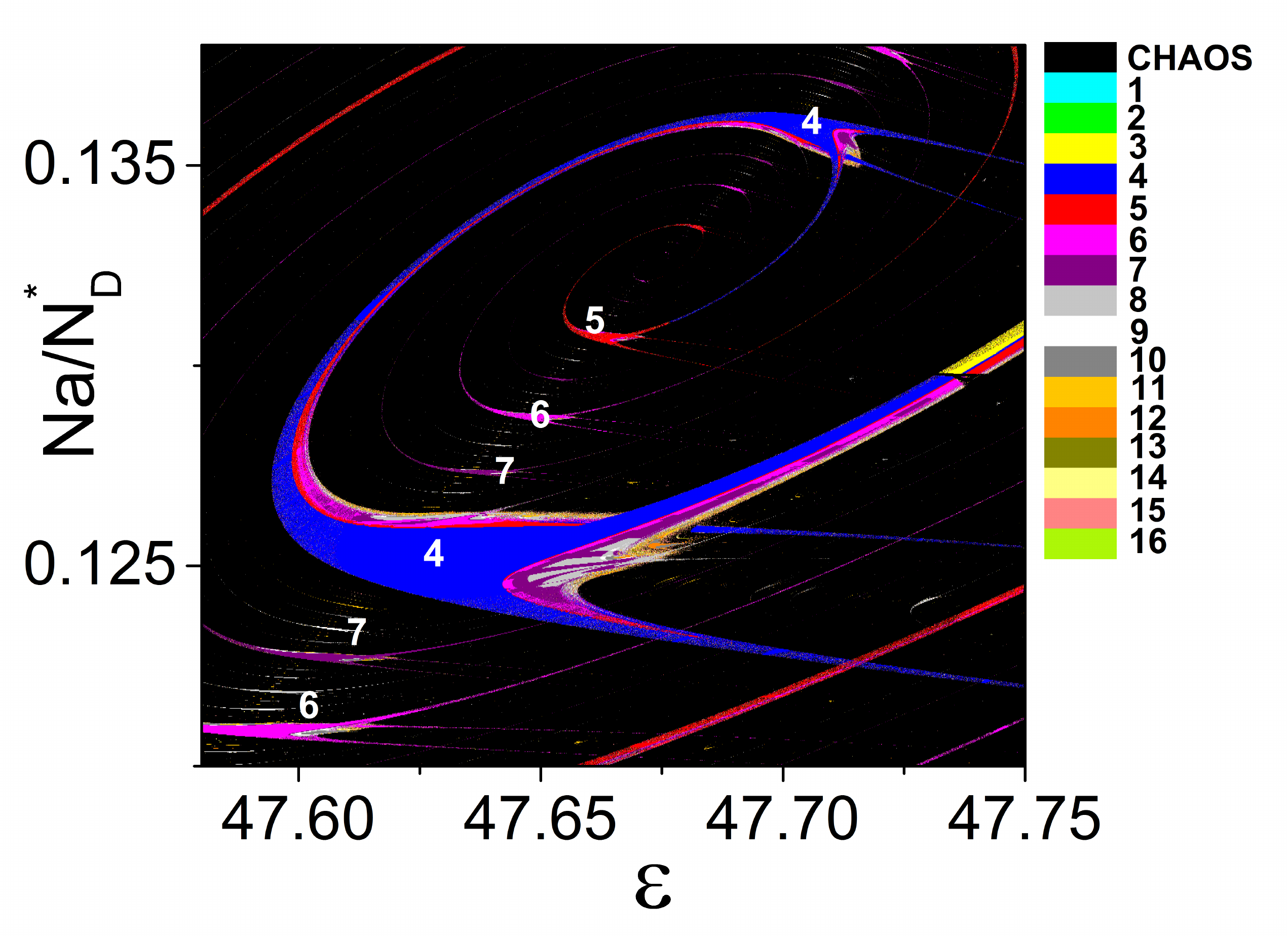}
\end{center}
\caption{(Color online) P-parameter-space for the SI-GaAs in resolution $8500\times900$ zoom inside Fig.7. Here, compensation factor $(N_A/N_D^*)$ was varied from $12.0\times10^{-2}$ up to $13.8\times10^{-2}$ with $2.0\times10^{-5}$ step and the electric field $(\epsilon)$ from 47.58 up to 47.75 with $2.0\times10^{-5}$ step.}
\label{Figure8}
\end{figure}

It was observed in the parameter space of Fig.~\ref{Figure7}, that one shrimp can connect with others in three ways simultaneously: The first way occurs in the antennas intersection between shrimps of different periodicities, the second way   with periodicity in sequence (P-3 to P-4, P-4 to P-5 and, so on) continuously connected by their antennas; and finally, the third connection is in the continuity between the ``tails" of shrimps with the same periodicity. The easy access between the self-organized structures constitutes a direct communication between the ``windows of order" within the ``chaos regions", producing news routes of bifurcation, thereby establishing a control of chaos. In Fig.~\ref{Figure8}, we present P-parameter-space for the SI-GaAs in resolution $(8500\times900)$ that is a zoom of Fig.~\ref{Figure7}.

A direct look of the ``shrimps" of Fig.~\ref{Figure7}, we can see that we have a big P-3 (yellow-shrimp), near ($\epsilon$ = 47.80) and $(N_A/N_D^*)$ = 0.14.  In the right, we have P-5 (red-shrimp) and P-6 (pink-shrimp) which is a period-adding sequence for $\epsilon$ ${>}$ 47.80. But in the left, we have P-4 (blue-shrimp) and P-5 (red-shrimp) which is a period-decreasing sequence, for $\epsilon$ ${<}$ 47.80. This result is similar, in essence, to the sequence obtained in \cite{Hoff14}. 

Comparing the experimental results \cite{DaSilvaRL04, Rubinger03, Albuquerque03} with numerical results obtained from our model, we can conclude that the ``two-valley model" successfully produces NNDC regions, and the chaotic behavior of the transport in SI-GaAs, which are characteristic of the nonlinear phenomena ``field enhance trapping" and ``impact ionization". The introduction of the $\Gamma$ and L valleys as two conduction channels with its characteristic drift velocity for free electrons were responsible for the success of our model.

\section{Conclusions}

In the present work, we proposed a physical model which we termed the ``two-valley model" which describes electrical conduction in semi-insulating GaAs (SI-GaAs). As far as we know, the proposed model is the first that, using generation recombination (g-r) rate equations, produces the negative differential conductivity regime in the N-shaped form (NNDC), which is the experimental signature of the nonlinear {J($\epsilon$)} characteristic curves of this material \cite{DaSilvaRL04, Rubinger03, Albuquerque03}. 
	
	In order to achieve this result, we have considered conduction channels in two valleys, namely, the $\Gamma$ and L valleys inside the conduction band and the fundamental state of the deep level defect. Besides that, it was necessary to determine an analytical expression for the drift velocity of the free-charges, by simulating the well-established experimental curve, available in the literature \cite{Ruch68}.
	
	Among the multiple alternative forms for g-r processes involving a deep-level defect and the two valleys, the proposed ``two-valley  model" was built as the smallest set tested that produced the NNDC region, and take into account the main SI-GaAs physical properties, as the ``impact ionization" and the ``field enhanced trapping" phenomena. With a small set of differential equations the computational cost is significantly reduced, on the evaluation of the nonlinear dynamics and on the time-series analysis, so the efficiency of the presented model increases.
	
	The nonlinear dynamics were investigated in the negative differential conductivity phenomena region in the N-shaped form (NNDC). Attractors were built by a combination of the time series of the defect density and the electric field, and many periodic and chaotic regimes were observed. A period-doubling bifurcation route-to-chaos with odd-periodic windows embedded was also observed in the bifurcation diagram of the maximum value of the electric field $\epsilon_{max}(t)$ .
	
	Evaluating a large quantity of time-series (up to $10^8$) for the presented system, and using a Periodicity-Detection (PD) routine, we were able to build the periodicity parameter-space for the SI-GaAs, which allowed the observation of self-organized periodic structures embedded in the chaotic regions, a ``shrimp" shaped in a spiral form, that forms a ``snail shell". This structure established a direct communication between the windows in order within chaotic regions, producing news routes of bifurcation. These intricate self-organized patterns were experimentally observed in the literature \cite{DaSilvaRL04, Rubinger03, Albuquerque03,DaSilvaRL06,DaSilvaSL09}, but still needed a full theoretical description of how they emerge, what their size in the parameter-space, if they have a fractal structure, based on the new model presented.
	
	The knowledge of detailed information on the parameter spaces is crucial to localize wide regions of smooth and continuous chaos, in order to fulfill the requirements for secure communications with chaos, and to explain the behavior of the nonlinear system when simultaneously two or more control parameters are changed.

\section{Acknowledgments}

Thanks CNPq, CAPES, FAPEMIG and FAPERJ, Brazilian official agencies for funding this work.


\end{document}